\documentclass[10pt,aps,prb,showpacs,floatfix,floats,superscriptaddress,twocolumn]{revtex4-1}
\usepackage{graphicx,latexsym}
\usepackage{dcolumn}
\usepackage{amssymb,amsmath,bm}
\usepackage{color}
\usepackage[normalem]{ulem}

\usepackage{graphicx}
\usepackage{dcolumn}
\usepackage[mathlines]{lineno}
\usepackage{amsthm,bm}
\usepackage{mathrsfs}
\DeclareMathAlphabet{\mathscrbf}{OMS}{mdugm}{b}{n}
\usepackage{etoolbox}
\usepackage{mathtools}
\usepackage{xparse}
\NewDocumentCommand \BiG {m O{} O{}}{
        \IfNoValueTF{ #2}
        {
                \IfNoValueTF{ #3}
                {
                        \big{.#1}
                }
                {
                        \big{.#1}^{ #3 }
                }
        }
        {
                \IfNoValueTF{ #3}
                {

                        \big{.#1}_{ #2}
                }
                {
                        \big{.#1}_{ #2}^{ #3}
                }
        }
}

\begin{document}

\title{Current correlations for the transport of interacting electrons\\ through
       parallel quantum dots in a photon cavity}

\author{Vidar Gudmundsson}
\email{vidar@hi.is}
\affiliation{Science Institute, University of Iceland, Dunhaga 3, IS-107 Reykjavik, Iceland}
\author{Nzar Rauf Abdullah}
\affiliation{Physics Department, College of Science, 
             University of Sulaimani, Kurdistan Region, Iraq}
\author{Anna Sitek}
\affiliation{Science Institute, University of Iceland, Dunhaga 3, IS-107 Reykjavik, Iceland}
\affiliation{Department of Theoretical Physics, Wroc{\l}aw University of Science and Technology, 50-370 Wroc{\l}aw, Poland}
\author{Hsi-Sheng Goan}
\email{goan@phys.ntu.edu.tw}
\affiliation{Department of Physics and Center for Theoretical Sciences, National Taiwan University, 
             Taipei 10617, Taiwan}
\affiliation{Center for Quantum Science and Engineering, 
             National Taiwan University, Taipei 10617, Taiwan}
\author{Chi-Shung Tang}
\email{cstang@nuu.edu.tw}
\affiliation{Department of Mechanical Engineering, National United University, Miaoli 36003, Taiwan}
\author{Andrei Manolescu}
\email{manoles@ru.is}
\affiliation{School of Science and Engineering, Reykjavik University, Menntavegur 
             1, IS-101 Reykjavik, Iceland}

%

\begin{abstract}
We calculate the current correlations for the steady-state electron transport through
multi-level parallel quantum dots embedded in a short quantum wire, that is placed in
a non-perfect photon cavity. We account for the electron-electron Coulomb interaction, and the 
para- and diamagnetic electron-photon interactions with a stepwise scheme of configuration
interactions and truncation of the many-body Fock spaces. In the spectral density of
the temporal current-current correlations we identify all the transitions, radiative and non-radiative,
active in the system in order to maintain the steady state. We observe strong signs of 
two types of Rabi oscillations.
\end{abstract}

\maketitle
%
%

\section{Introduction}
Experiments\cite{PhysRevX.6.021014,Delbecq11:01,2017arXiv170401961L,PhysRevX.7.011030,Frey11:01,Mi}
in which the electron transport through nanoscale electronic systems placed
in photon cavities, and model calculations\cite{PhysRevLett.116.113601,PhysRevB.92.165403,Gudmundsson16:AdP_10,2017arXiv170300803H} 
thereof, are gaining attention in the last years.

Due to small size of the electronic systems the constant average current through the system in 
the steady state does not convey much information about the underlying processes,
and one might expect information about radiative transitions to be lost at that time scale,
or not detectable.\cite{2016arXiv161109453G} In order to remedy this situation researchers have realized
that the noise power spectrum, or the noise power spectral density of a system calculated through the 
Fourier transform of the current-current two-time correlation function can be measured
experimentally.\cite{0957-4484-21-27-272001} Many theoretical researchers have used this to
calculate the noise spectral density for electron transport through model systems in different
situations using, for example, non-equilibrium Green functions,\cite{PhysRevB.43.4534} Markovian master 
equation in the steady state,\cite{PhysRevB.59.10748} or non-Markovian master equations in the transient 
regime,\cite{PhysRevB.89.115411} just to mention very few.

Complementary to the calculation of the noise power spectral densities of the charge current transport
through electron systems on the nanoscale, the calculation of the power spectral properties 
of photon emission statistics of cavities with embedded electron systems has been undertaken by many 
more theoretical groups.\cite{PhysRevA.80.053810,PhysRevA.74.033811,Ciuti05:115303,PhysRevLett.116.113601} 
Recently, we have investigated the photon correlations in the emission radiation from a photon cavity
containing a short quantum wire with embedded two parallel quantum dots through which a steady state
current is driven with a bias difference between two external leads.\cite{2017arXiv170603483G}
There, the spectral density of the fluctuations in the radiation can be used to differentiate between the
the conventional and the ground state electroluminescence in the strong electron-photon coupling
regime.\cite{PhysRevLett.116.113601,2017arXiv170603483G} Here, we will demonstrate that in this complex
interacting many-state system, the power spectral density of the temporal current-current correlations can be used to identify the 
underlying processes, the transitions between interacting many-body states of cavity-photon dressed 
electron states, that contribute to maintaining the system in its steady state.

\section{Model}
\begin{figure}[htb]
      \includegraphics[width=0.48\textwidth]{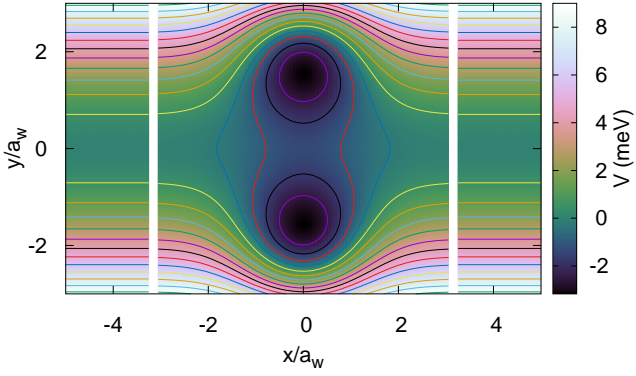}
      \caption{The potential energy landscape defining the parallel quantum dots 
               embedded in a short quantum wire of length $150$ nm $\approx$ 6.3$a_w$,
               where $a_w=23.8$ nm is the effective magnetic length for magnetic field
               $B=0.1$ T and parabolic confinement energy $\hbar\Omega_0=2.0$ meV of the 
               short wire and leads in the $y$-direction. The gaps at $x\approx\pm{3.15}a_w$
               indicate the onset of the semi-infinite leads.}
      \label{System}
\end{figure}

We consider a short two-dimensional GaAs quantum wire with length $L=150$ nm
placed in a photon cavity. We use the 36 lowest in energy single-electron states
of the wire, $|i\rangle$, to build a many-electron Fock space of 0-3 Coulomb interacting 
electrons, $|\mu )$. The potential defining the short quantum wire with two parallel 
quantum dots displayed in Fig.\ \ref{System} is
\begin{align}
      V(x,y) =& \left[\frac{1}{2}m^*\Omega_0^2y^2 +eV_g\right.\nonumber\\
             +& \left. V_d\sum_{i=1}^2\exp{\left\{-(\beta x)^2+\beta^2(y-d_i)^2\right\}} \right]\nonumber\\
             \times&\theta\left(\frac{L_x}{2}-|x|\right)
\label{Potential}
\end{align}
with $\hbar\Omega_0 = 2.0$ meV, $V_d = -6.5$ meV, $\beta = 0.03$ nm$^{-1}$, 
$d_1=-50$ nm, $d_2=+50$ nm, $L_x = 150$ nm, and $\theta$ is the Heaviside 
step function. The plunger gate voltage $V_g$ is used to move the states of the
system up or down with respect to the bias window defined by the external leads to
be describe below.

We use as a kernel for the mutual electron-electron Coulomb interaction
\begin{equation}
      V_{\mathrm{Coul}}(\mathbf{r}-\mathbf{r}') = \frac{e^2}{\kappa_e\sqrt{|\mathbf{r}-\mathbf{r}'|^2+\eta_c^2}},
\end{equation}
with a small regularizing parameter $\eta_c/a_w=3\times 10^{-7}$ ($a_w$ being defined below), 
and for GaAs parameters we assume $\kappa_e =12.4$, $m^*=0.067m_e$, and $g^*=-0.44$. 
In terms of field operators the Hamitonian of the central system is
\begin{align}
      H_\mathrm{S} =& \int d^2r \psi^\dagger (\mathbf{r})\left\{\frac{\pi^2}{2m^*}+
        V(\mathbf{r})\right\}\psi (\mathbf{r})
        + H_\mathrm{EM} + H_\mathrm{Coul}\nonumber\\ 
        -&\frac{1}{c}\int d^2r\;\mathbf{j}(\mathbf{r})\cdot\mathbf{A}_\gamma
        -\frac{e}{2m^*c^2}\int d^2r\;\rho(\mathbf{r}) A_\gamma^2,
\label{Hclosed}
\end{align}
with 
\begin{equation}
      {\bm{\pi}}=\left(\mathbf{p}+\frac{e}{c}\mathbf{A}_{\mathrm{ext}}\right),
\end{equation}
where $\mathbf{A}_{\mathrm{ext}}$ is a classical vector potential producing an external homogeneous
small magnetic field $B=0.1$ T along the $z$-axis, perpendicular to the plane of the two-dimensional 
quantum wire, inserted to break the spin and the orbital degeneracies of the states in order to enhance 
the stability of the results. The first term in the second line of Eq.\ (\ref{Hclosed}) is the paramagnetic,
and the second term the diamagnetic, electron-photon interaction. 
The external magnetic field, $B$, and the parabolic confinement energy of the leads and the 
central system $\hbar\Omega_0=2.0$ meV, together with the cyclotron frequency 
$\omega_c=(eB)/(m^*c)$ lead to an effective characteristic confinement energy 
$\hbar\Omega_w=\hbar({\omega_c^2+\Omega_0^2})^{1/2}$, and an effective magnetic length 
$a_w=(\hbar /(m^*\Omega_w))^{1/2}$.
This characteristic length scale assumes approximately the value 23.8 nm for the parameters selected here.
In terms of the cavity photon creation and annihilation 
operators, $a^\dagger$ and $a$, the Hamiltonian for the single cavity photon mode is 
$H_\mathrm{EM}=\hbar\omega a^\dagger a$, with energy $\hbar\omega$.

We assume a rectangular photon cavity $(x,y,z)\in\{[-a_\mathrm{c}/2,a_\mathrm{c}/2]
\times [-a_\mathrm{c}/2,a_\mathrm{c}/2]\times [-d_\mathrm{c}/2,d_\mathrm{c}/2]\}$ 
with the short quantum wire centered in the $z=0$ plane. In the Coulomb gauge the
polarization of the electric field parallel to the transport
in the $x$-direction (with the unit vector $\mathbf{e}_x$) is accomplished in the TE$_{011}$ mode, 
or perpendicular (defined by the unit vector $\mathbf{e}_y$) in the 
TE$_{101}$ mode. 
The two versions of the quantized vector potential for the cavity field are in a stacked notation 
expressed as
\begin{equation}
      \mathbf{A}_\gamma (\mathbf{r})=\left({\hat{\mathbf{e}}_x \atop \hat{\mathbf{e}}_y}\right)
      {\cal A}\left\{a+a^{\dagger}\right\}
      \left({\cos{\left(\frac{\pi y}{a_\mathrm{c}}\right)}\atop\cos{\left(\frac{\pi x}{a_\mathrm{c}}\right)}} \right)
      \cos{\left(\frac{\pi z}{d_\mathrm{c}}\right)},
\label{Cav-A}
\end{equation}
for the TE$_{011}$ and TE$_{101}$ modes, respectively. The strength of the vector potential, ${\cal A}$,
determines the coupling constant $g_\mathrm{EM} = e{\cal A}\Omega_wa_w/c$, here set to 0.05 meV,
or 0.10 meV, leaving a dimensionless polarization tensor
\begin{equation}
      g_{ij}^k = \frac{a_w}{2\hbar}\left\{\langle i|\hat{\mathbf{e}}_k\cdot\bm{\pi}|j\rangle + \mathrm{h.c.}\right\}.
\end{equation}

The coupling of the central system to the leads is described by the Hamiltonian
\begin{equation}
    H_T = \theta (t)\sum_{il} \int d  \mathbf{q} \left(T_{\mathbf{q}i}^l c_{\mathbf{q}l}^\dagger d_i +
          (T_{\mathbf{q}i}^l )^* d_i^\dagger c_{\mathbf{q}l}\right),
\label{H_T}
\end{equation} 
where $d_i$ is an annihilation operator for the single-electron state $|i\rangle$ of the 
central system, $c_{\mathbf{q}l}$ an annihilation operator for an electron in lead $l\in\{L,R\}$ 
in state $|{\mathbf q}\rangle$, with ${\mathbf q}$ standing for the momentum $q$ and the subband index
$n_l$ in the semi-infinite quasi-one dimensional lead. The coupling tensor $T_{\mathbf{q}i}^l$ depends
on the nonlocal overlap of the single-electron states at the internal boundaries in the central system and the respective 
lead.\cite{Gudmundsson09:113007,Moldoveanu09:073019,Gudmundsson12:1109.4728}
This setup is intended for a weak tunneling coupling of the central system with the leads, but allows
allows for full coupling between the quantum dots and the rest of the central system, like
in a scattering approach.\cite{Gudmundsson05:BT}
The remaining overall coupling constant to the leads is $g_{\mathrm{LR}}a_w^{3/2}=0.124$ meV,
in the weak coupling limit used here.

As we are interested in the properties of system in the steady state here, we transform a 
non-Markovian master equation built according to the projection formalism of Nakajima\cite{Nakajima58:948}
and Zwanzig\cite{Zwanzig60:1338} to a Markovian equation\cite{2016arXiv161003223J} for the reduced density operator 
of the central system
\begin{align}
      \partial_t\rho_\mathrm{S}(t) = &-\frac{i}{\hbar}[H_\mathrm{S},\rho_\mathrm{S}(t)]
      -\left\{\Lambda^L[\rho_\mathrm{S} ;t]+\Lambda^R[\rho_\mathrm{S} ;t]\right\}\nonumber\\
      &-\frac{\kappa}{2}(\bar{n}_R+1)\left\{2a\rho_\mathrm{S}a^\dagger - a^\dagger a\rho_\mathrm{S} - \rho_\mathrm{S}a^\dagger a\right\}\nonumber\\
       &-\frac{\kappa}{2}(\bar{n}_R)\left\{2a^\dagger\rho_\mathrm{S}a - aa^\dagger\rho_\mathrm{S} - \rho_\mathrm{S}aa^\dagger\right\} ,
\label{NZ-eq}
\end{align}
where the last two terms in the first line describe the ``dissipation'' caused by the Left and Right leads.
The dissipation terms are constructed with terms up to second order in the coupling Hamiltonian
(\ref{H_T}), but without resorting to the rotating wave approximation, as more than one resonance 
with the photon field can be active to some extent in the system for each set of parameters used in 
the calculations. The dissipation terms in Eq.\ (\ref{NZ-eq}) are\cite{2016arXiv161003223J}
\begin{equation}
    \Lambda^l[\rho_\mathrm{S} ;t]= \dfrac{1}{\hbar^2}\int d\epsilon D^l(\epsilon)\theta(t) \left\lbrace \left[ \tau^l, \Omega[\rho_\mathrm{S}]
        \right] + \mathrm{h.c.} \right\rbrace
\end{equation}
with $\theta (t)$ the Heaviside unit step function, $\tau^l$ the many-body version of the coupling tensor of lead $l$, and
\begin{equation} 
    \BiG{\Omega}[\alpha \beta][]\left[\rho_\mathrm{S}\right] = \left\lbrace   \BiG{\mathcal{R}[\rho_\mathrm{S} (t)]}[\alpha \beta][]- 
    \BiG{\mathcal{S}[\rho_\mathrm{S} (t)]}[\alpha \beta][]
    \right\rbrace \BiG{\delta}[][\beta \alpha],
\end{equation}
where 
\begin{equation}
        \BiG{\delta}[][\beta \alpha] = \BiG{\delta}\left(E_\beta - E_\alpha - \epsilon\right).
\end{equation}
The density of states in leads $l$ is $D^l(\epsilon) = |{d\mathbf{q}}/{d\epsilon}|$, and 
we have defined the superoperators
\begin{align} \label{eq:SuperoperatorsRS}
    \mathcal{S}[\rho_\mathrm{S}] &=  S\rho_\mathrm{S} ,\\
    \mathcal{R}[\rho_\mathrm{S}] &= \rho_\mathrm{S} R,
\end{align}
from
\begin{align}
    R &= \pi (1 - F^l)(\tau^l)^\dagger
    \intertext{and}
    S &= \pi F^l (\tau^l)^\dagger ,
\end{align}
with $F^l$ being the equilibrium Fermi distribution in lead $l$.
The last two lines of Eq.\ (\ref{NZ-eq}) describe the Markovian photon decay of a non-perfect cavity 
with an overall decay constant, $\kappa$, and a mean value of photons in the reservoir $\bar{n}_R$.
The non-interacting electron gas in the leads is at temperature $T=0.5$ K, corresponding 
to the thermal energy $k_BT\approx 0.043$ meV.

The charge and the charge-current density operators of the central system are 
\begin{equation}
      \rho       = -e\psi^\dagger\psi, \quad
      \mathbf{j} = -\frac{e}{2m^*}\left\{\psi^\dagger\left({\bm{\pi}}\psi\right)
                 +\left({\bm{\pi}}^*\psi^\dagger\right)\psi\right\}.
\end{equation}
Due to the structure of the master equation (\ref{NZ-eq}) the time-dependent average current from the left lead into the central 
system, and the current from it into the right lead can be calculated as
\begin{equation}
      I_l(t) = \mathrm{Tr}_\mathrm{S}\left\{ \Lambda^l[\rho_\mathrm{S};t]Q\right\}, l\in\{L,R\}, 
\label{Il-mean}
\end{equation}
where $Q=-e\sum_id^\dagger d_i$ is the charge operator of the central system. The current-current
correlation functions are best written for the corresponding operators in the Heisenberg picture
\begin{equation}
      D_{ll'}(\tau ) = \langle I_l(\tau )I_{l'}(0)\rangle ,\quad \tau > 0,
\label{Dll}
\end{equation}
and for a calculation of it in the steady state we redefine the time point $t=0$ to 
refer to any time at which the system has reached its steady state. In the time domain
a more convenient correlation function is
\begin{equation}
      S_{ll'}(\tau ) = \langle \delta I_l(\tau )\delta I_{l'}(0)\rangle /I(0)^2,
\end{equation}
where $\delta I_l(\tau ) = I_l(\tau )-\langle I_l(\tau ) \rangle$, and the two functions are related via
\begin{equation}
      S_{ll'}(\tau ) = D_ {ll'}(\tau )/I(0)^2 -1,
\end{equation}
as in the steady state $\langle I_l(\tau )\rangle = \langle I_{l'}(0)\rangle = I(0)$.

Despite the simple look of Eq.\ (\ref{Il-mean}) one realizes that the construction of a current operator
is not straight forward having in mind that the dissipation terms, $\Lambda^l[\rho_\mathrm{S};t]$, have
the reduced density operator to the left, the right, or sandwiched between system operators. The solution
is to use the concept of superoperators (of which the Liouville operator is one) or go one step further
and use a Liouville space representation.\cite{Weidlich71:325,Haake73:98,Nakano2010}
We take the latter option and express the mean value of the current as
\begin{equation}
      I_l(t) = \mathrm{Tr}_\mathrm{S}\left\{ Q\left[ \Lambda^l\left( \rho_\mathrm{S}(t)\right)_\mathrm{vec}\right]_\mathrm{Mat}\right\}, 
\label{IlL-mean}
\end{equation}
where $\Lambda^l$ is a $N_\mathrm{mes}^2\times N_\mathrm{mes}^2$ dimensional matrix in Liouville space representing the dissipation,
the ``vec'' operation stacks the $N_\mathrm{mes}$ columns of the matrix representing $\rho_\mathrm{S}$ in the Fock space into a vector
in Liouville space, 
and the ``Mat'' operation reverses that procedure. $N_\mathrm{mes}=120$ is the number of many-body states in our Fock basis of 
cavity-photon dressed electron states. 


Expression (\ref{IlL-mean}) suggests using $Q\Lambda^l\cdot$ as the current operator 
and the Quantum Regression Theorem (QRT)\cite{0305-4470-14-10-013,Wallis-QO} that is valid in the Markovian 
limit for weak system-leads coupling.\cite{doi:10.1063/1.3570581,PhysRevA.47.1652,PhysRevB.59.10748,PhysRevB.64.235307,PhysRevB.92.165403}
The QRT states that the equation of motion for the two-time correlation function is of the
same form as the Markovian master equation for the reduced density operator of the system, 
but for an effective density operator,\cite{doi:10.1063/1.3570581} which for the current correlation is 
\begin{equation}
      \chi^l (\tau ) = \mathrm{Tr}_\mathrm{R}\left\{ e^{-iH\tau /\hbar}Q\Lambda^l\rho(0)e^{+iH\tau /\hbar}\right\} ,
\end{equation}
with $H$ the Hamiltonian of the total system, $\rho (0)$ its density operator after the on-set of the steady state,
redefinig that point of time to be $t=0$. 
$\mathrm{Tr}_\mathrm{R}$ is the trace operator with respect to the variables of the reservoir.
The two-time average or the correlation function is then
\begin{equation}
      D_{ll'}(\tau )=\langle I_l (\tau ) I_{l'}(0)\rangle = \mathrm{Tr}_\mathrm{S}\left\{ I_{l'}(0) \chi^l (\tau ) \right\} , 
\end{equation}
where  $\mathrm{Tr}_\mathrm{S}$ is the trace operation with respect to the state space of the central system.


\section{Results for one electron ground state at $\mathbf{V_g=2.0}\ \mathbf{mV}$}
\begin{figure}[htb]
      \includegraphics[width=0.48\textwidth]{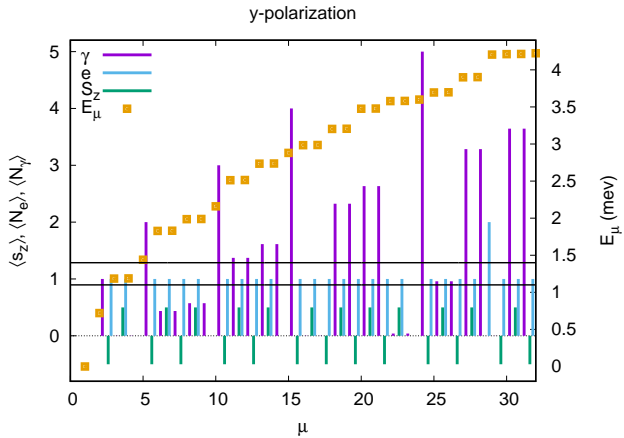}
      \caption{The properties of the 32 lowest in energy many-body states for $V_g=2.0$ mV
               and $y$-polarized cavity photon field.
               The horizontal black lines represent
               the chemical potentials of the left lead $\mu_L=1.40$ meV, the right lead $\mu_R=1.10$ meV, 
               $\hbar\omega = 0.72$ meV, and $g_\mathrm{EM}=0.05$ meV.
               The squares indicate the energy $E_\mu$ of each state $|\breve{\mu})$, 
               and the impulses show the photon expectation value (labeled with $\gamma$), the electron number 
               (labeled with $e$), and the $z$-component of the spin ($S_z$).}
      \label{NeNphESz-Vgp2p00-hw0p72-muL1p40-gEM0p20-y}
\end{figure}

We use two cases for different values of the plunger gate voltage $V_g$
to show how Rabi resonances influence and turn up in the 
steady state properties of the system in different ways. 
We select a rather narrow bias window with $\mu_L=1.4$ meV, and $\mu_L=1.1$ meV,
and investigate the current-current correlations for two different cases in the steady state.
For $V_g=+2.0$ mV, when only the two spin components of the one-electron ground state are
within the bias window. In this case the photon energy is selected to be $\hbar\omega =0.72$ meV
to promote a Rabi resonance between the one-electron ground state and the first excitation thereof.
The properties of the 32 lowest in energy many-body states of the system are displayed in
Fig.\ \ref{NeNphESz-Vgp2p00-hw0p72-muL1p40-gEM0p20-y} for the case of a $y$-polarized photon field.

As we have discussed earlier, the symmetry properties of the states of parallel quantum dots
lead to a large Rabi resonance for the $y$-polarization caused by the paramagnetic 
electron-photon interaction, but a very small resonance for the $x$-polarized field that is
only caused by the diamagnetic part of the electron-photon interaction.\cite{2016arXiv161109453G}
\begin{figure}[htb]
      \includegraphics[width=0.48\textwidth]{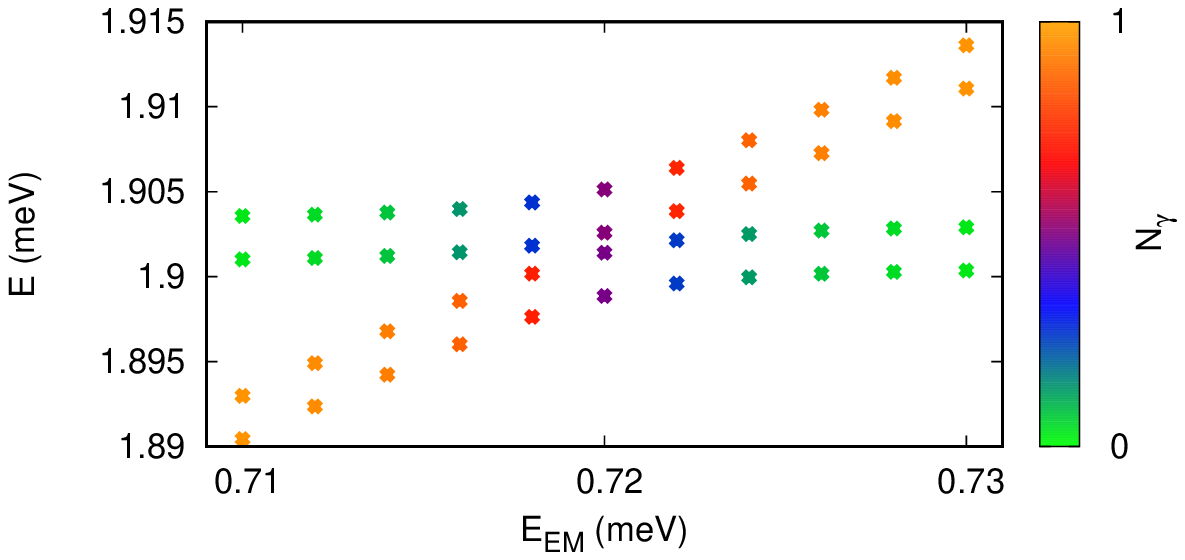}
      \includegraphics[width=0.48\textwidth]{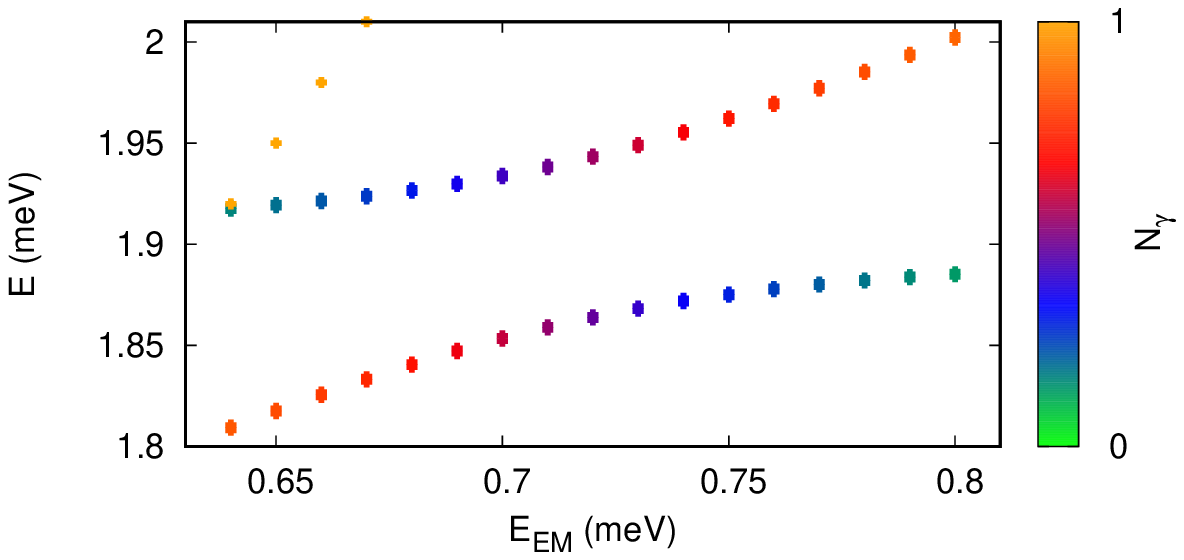}
      \caption{The Rabi-splitting of the two spin components of the 
               first excitation of the one-electron ground state 
               as a function of the photon energy $E_\mathrm{EM}=\hbar\omega$ for $V_g=2.0$ mV 
               and $x$- (upper), and $y$-polarized photon field (lower).
               $g_\mathrm{EM}=0.05$ meV.}
      \label{Rabi-072}
\end{figure}
These two resonances are shown in Fig.\ \ref{Rabi-072}.

\begin{figure}[htb]
      \includegraphics[width=0.48\textwidth]{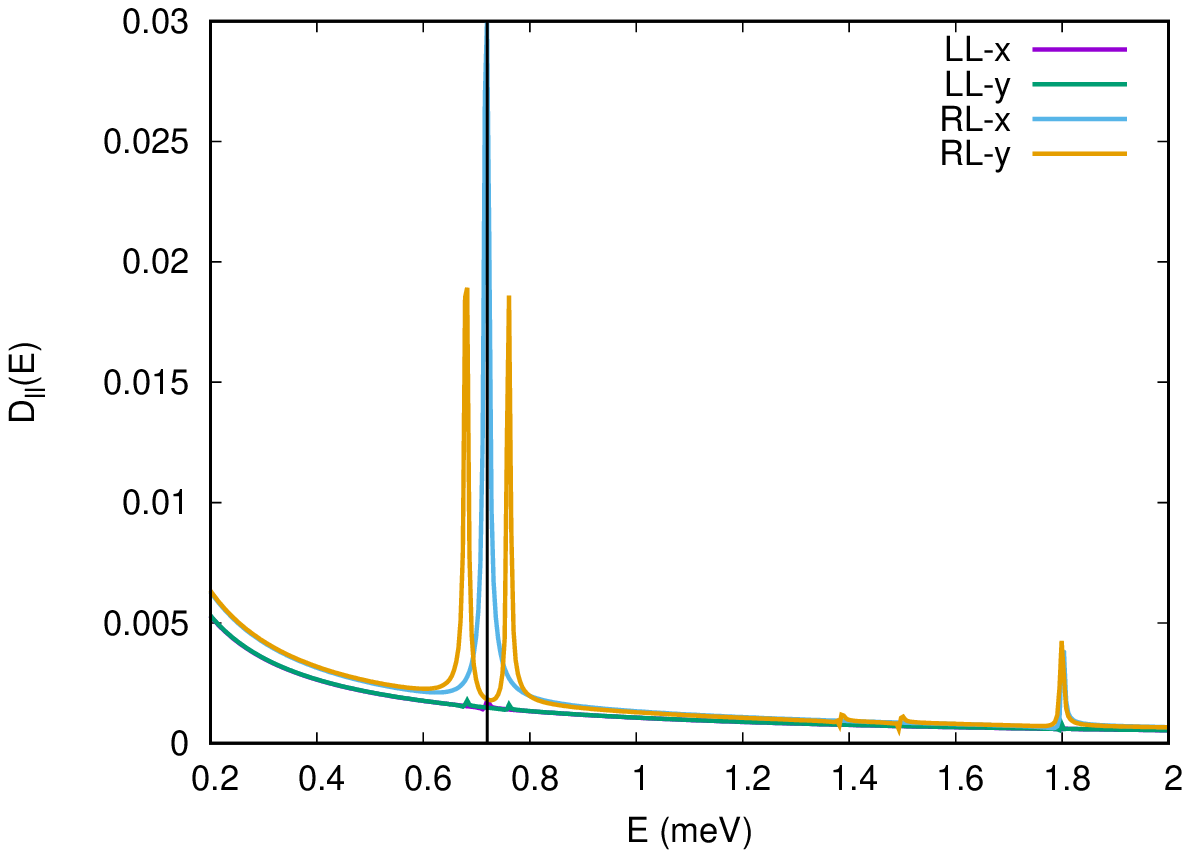}
      \includegraphics[width=0.48\textwidth]{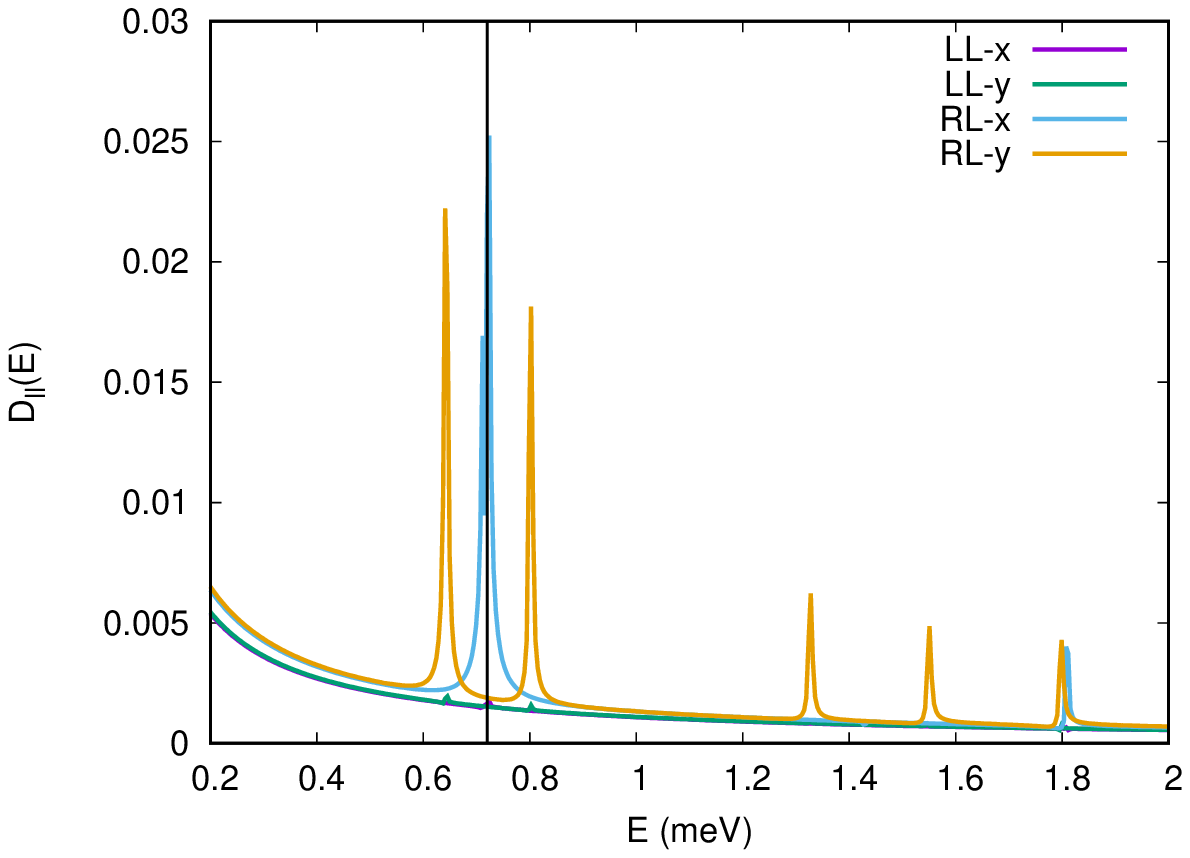}
      \caption{The spectral densities for the current-current correlations for $g_\mathrm{EM}=0.05$ meV (upper),
               and $g_\mathrm{EM}=0.10$ meV (lower). $V_g=2.00$ mV, and $\hbar\omega=0.72$ meV 
               (marked by a thin black vertical line).
               The key for the curves gives $l$ and $l'$ in terms of $L$ and $R$, and the polarization
               of the photon field. $\kappa = 10^{-3}$ meV.}
      \label{fft-LL-RL-im-0G-Vgp2p00-dis1p0Em03-muL1p40-nrr0-hw0p72-gEM0px0}
\end{figure}

The spectral densities of the current-current correlations, $D_{ll'}(E)$, are displayed in 
Fig.\ \ref{fft-LL-RL-im-0G-Vgp2p00-dis1p0Em03-muL1p40-nrr0-hw0p72-gEM0px0}, and the identity
of the main peaks for the case of $y$-polarized photons (in the lower panel of 
Fig.\ \ref{fft-LL-RL-im-0G-Vgp2p00-dis1p0Em03-muL1p40-nrr0-hw0p72-gEM0px0}) is listed in Table \ref{Vg200-transitions}. 
Not surprisingly, the almost degenerate two spin components of the one-electron ground state
$|\breve{03})$ and $|\breve{04})$, the only states placed in the bias window,
are the initial states for all transitions. 
\begin{table}
\begin{tabular}{c|c}
Peak (meV)  &   Transitions \\
\hline
0.642       &   $|\breve{03})\leftrightarrow|\breve{06})$, $|\breve{04})\leftrightarrow|\breve{07})$\\
0.801       &   $|\breve{03})\leftrightarrow|\breve{08})$, $|\breve{04})\leftrightarrow|\breve{09})$\\
1.33        &   $|\breve{03})\leftrightarrow|\breve{11})$, $|\breve{04})\leftrightarrow|\breve{12})$\\
1.55        &   $|\breve{03})\leftrightarrow|\breve{13})$, $|\breve{04})\leftrightarrow|\breve{14})$\\
1.80        &   $|\breve{03})\leftrightarrow|\breve{16})$, $|\breve{04})\leftrightarrow|\breve{17})$\\
\hline
\end{tabular}
\caption{Identification of the peaks seen in the lower panel of Fig.\ \ref{fft-LL-RL-im-0G-Vgp2p00-dis1p0Em03-muL1p40-nrr0-hw0p72-gEM0px0}
         for the spectral density of the current-current correlations in the case of a $y$-polarized photon field. 
         $g_\mathrm{EM}=0.10$ meV, $V_g=2.00$ mV, $\hbar\omega=0.72$ meV, and $\kappa = 10^{-3}$ meV.}
\label{Vg200-transitions}
\end{table}
The first two lines in Table \ref{Vg200-transitions} refer to transitions from both spin components
of the one-electron ground state, $|\breve{03})$ and $|\breve{04})$, to the Rabi-split first excitation
thereof, \{$|\breve{06})=R^-_\downarrow$, $|\breve{07})=R^-_\uparrow$\}, and \{$|\breve{08})=R^+_\downarrow$, 
$|\breve{09})=R^+_\uparrow$\}, for photon energy $\hbar\omega=0.72$ meV.

For $V_g=2.0$ mV there are no electronic states of the central system below the bias window, and the two next lines
in Table \ref{Vg200-transitions} identify transitions to higher order states of the Rabi resonance, in the
sense that the pairs $\{|\breve{11})$, $|\breve{12})\}$, and  $\{|\breve{13})$, $|\breve{14})\}$
have a mean photon number in the range 1 to 2. The last line in Table \ref{Vg200-transitions} is for the
last peak easily visible in the lower panel of Fig.\ \ref{fft-LL-RL-im-0G-Vgp2p00-dis1p0Em03-muL1p40-nrr0-hw0p72-gEM0px0} 
which is caused by a transition to the states $|\breve{16})$, and $|\breve{17})$, that only have a very small
photon component. This last fact conforms to that the last peak has the same size and location in the 
upper and lower panel of Fig.\ \ref{fft-LL-RL-im-0G-Vgp2p00-dis1p0Em03-muL1p40-nrr0-hw0p72-gEM0px0}, i.e.\
it is independent of the electron-photon coupling strength, $g_\mathrm{EM}$.

An important point to notice is that the energy distance of the first two peaks reflects directly the 
Rabi-splitting as the electron-photon coupling is increased from the upper to the lower panel
in Fig.\ \ref{fft-LL-RL-im-0G-Vgp2p00-dis1p0Em03-muL1p40-nrr0-hw0p72-gEM0px0}. The spectral density of the photon-photon 
correlation reveals three peaks,\cite{2017arXiv170603483G} the so-called Mollow 
triplet,\cite{PhysRev.188.1969} but, here, in the spectral density of the current-current correlations
there are only two peaks.

For an $x$-polarized cavity photons the Rabi splitting is much smaller, of the same order as the spin
splitting in GaAs for $B=0.1$ T, and a careful inspection of the data shows the Rabi resonance peak starting to split
into two parts.

\section{Results for two-electron ground state at $\mathbf{V_g=0.5}\ \mathbf{mV}$}
Now, we turn to a very different case in our system, by reducing the plunger gate voltage to
$V_g=0.5$ mV. With the same bias window as before, we have only the two-electron ground state
within it, a singlet, and might expect similar phenomena taking place as when only the one-electron ground state
was within the bias window. Before analyzing the results, we remind the reader that there are one-electron
states below the bias window, and in the weak coupling limit with only sequential tunneling between the leads and the
central system the current through a two-electron state is very low. A third very important fact is that in a multi-state
system, even though the photon energy is tuned close to a certain resonance, there can always be other weaker, more
detuned, resonances at play in the system. This last fact is also a very good reason to include both the para- and
the diamagnetic electron-photon interactions in the model.

\begin{figure}[htb]
      \includegraphics[width=0.48\textwidth]{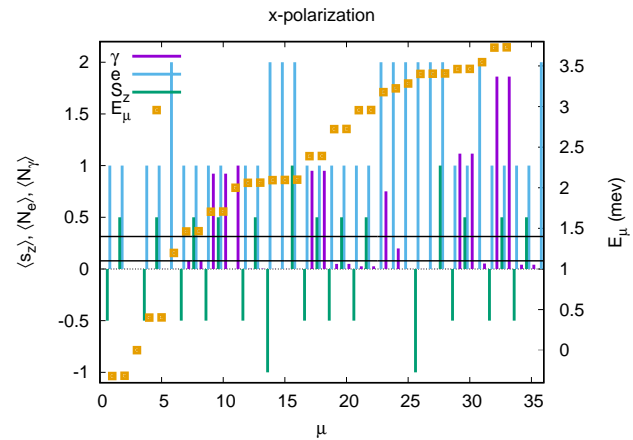}
      \caption{The properties of the 36 lowest in energy many-body states for $V_g=2.0$ mV
               and $x$-polarized cavity photon field.
               The horizontal black lines represent
               the chemical potentials of the left lead $\mu_L=1.40$ meV, the right lead $\mu_R=1.10$ meV, 
               $\hbar\omega = 2.00$ meV, and $g_\mathrm{EM}=0.05$ meV.
               The squares indicate the energy $E_\mu$ of each state $|\breve{\mu})$, 
               and the impulses show the photon expectation value (labeled with $\gamma$), the electron number 
               (labeled with $e$), and the $z$-component of the spin ($S_z$).}
      \label{NeNphESz-Vgp0p50-hw2p00-muL1p40-gEM0p10-x}
\end{figure}

In order to analyze the results, Fig.\ \ref{NeNphESz-Vgp0p50-hw2p00-muL1p40-gEM0p10-x} displays
the properties of the 36 lowest in energy many-body eigenstates of the closed central system
for the case of an $x$-polarized cavity photon field. It is proper here to remind the reader that the 
numbering of the photon-dressed electron states changes as the plunger gate voltage is changed.

\begin{figure}[htb]
      \includegraphics[width=0.48\textwidth]{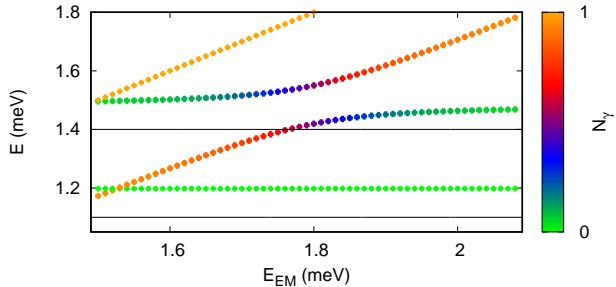}
      \caption{The Rabi-splitting of the two spin components of the 
               of the one-electron states $|\breve{07})$ and $|\breve{09})$ (just above the bias window for $E_\mathrm{EM}>1.8$ meV)
               as a function of the photon energy $E_\mathrm{EM}=\hbar\omega$, $V_g=0.5$ mV, and 
               an $x$-polarized photon field. The two-electron ground state is seen with a horizontal dispersion close to
               $E\approx 1.2$ meV. $g_\mathrm{EM}=0.05$ meV. The bias window is indicated with two thin horizontal black lines.}
      \label{Rabi-200}
\end{figure}

The plunger gate voltage is set at $V_g=0.5$ mV, and the two-electron ground state $|\breve{06})$ is coupled to the
first excitation thereof by selecting the photon energy $\hbar\omega =2.0$ meV, resulting in the Rabi split states
$|\breve{23})$ and $|\breve{24})$. Due to the low current through two-electron states,
the one-electron states just above the bias window, $|\breve{07})$ and $|\breve{08})$,
play a key role in the transport through the system. 
For the $x$-polarized cavity photon field these states are the 
lower energy states in Rabi split pairs with the states $|\breve{09})$ and $|\breve{10})$ as the upper states,
for photon energy $E_\mathrm{EM}=\hbar\omega\approx 1.8$ meV as can be seen in Fig.\ \ref{Rabi-200}. 
This splitting is only strong for the $x$-polarization as the underlying electronic states have odd parity in the $x$-direction, 
but even parity in the $y$-direction.\cite{2017arXiv170603483G} 

\begin{figure}[htb]
      \includegraphics[width=0.48\textwidth]{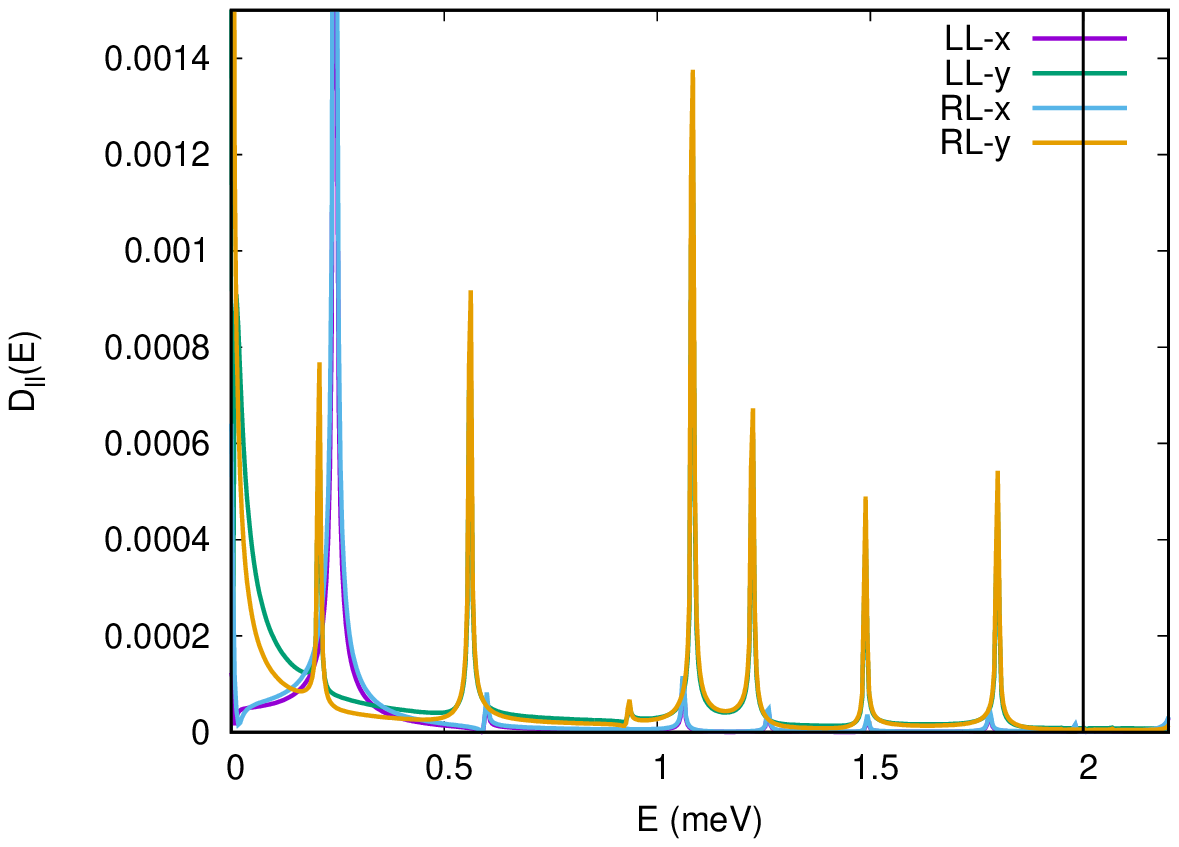}
      \includegraphics[width=0.48\textwidth]{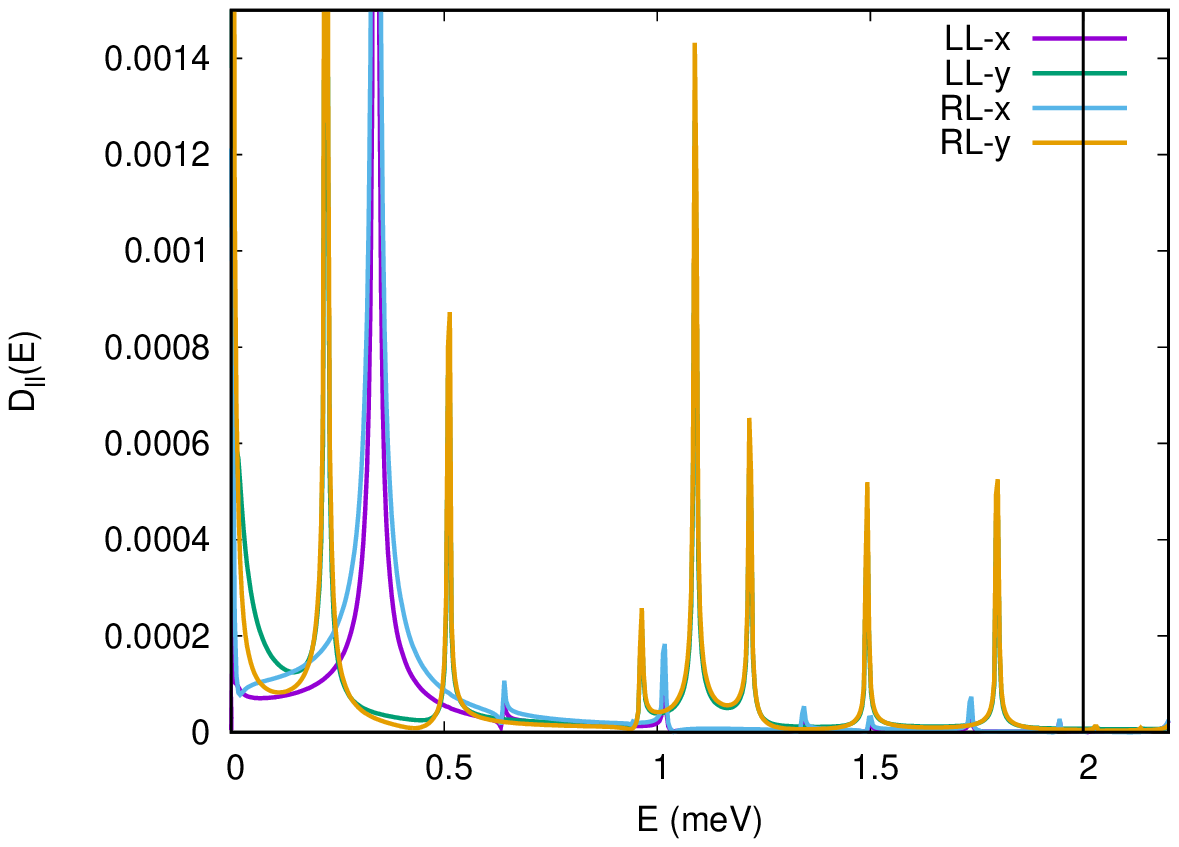}
      \caption{The spectral densities for the current-current correlations for $g_\mathrm{EM}=0.05$ meV (upper),
               and $g_\mathrm{EM}=0.10$ meV (lower). $V_g=0.50$ mV, $\hbar\omega=2.00$ meV (marked by a thin vertical line), 
               and $\kappa = 10^{-4}$ meV.}
      \label{fft-LL-RL-im-0G-Vgp0p50-dis1p0Em04-muL1p40-nrr0-hw2p00-gEM0px0}
\end{figure}

The spectral density of the current-current correlations, $D_{ll'}(E)$, is shown in 
Fig.\ \ref{fft-LL-RL-im-0G-Vgp0p50-dis1p0Em04-muL1p40-nrr0-hw2p00-gEM0px0},
and the peaks are correlated with transitions in the central system in Table \ref{Vg050-transitions}. 
\begin{table}
\begin{tabular}{c|c}
Peak (meV)  &   Transitions \\
\hline
0.204       &   $|\breve{07})\leftrightarrow|\breve{09})$, $|\breve{08})\leftrightarrow|\breve{10})$\\
0.561       &   $|\breve{07})\leftrightarrow|\breve{12})$, $|\breve{08})\leftrightarrow|\breve{13})$\\
0.932       &   $|\breve{07})\leftrightarrow|\breve{17})$, $|\breve{08})\leftrightarrow|\breve{18})$\\
1.082       &   $|\breve{07})\leftrightarrow|\breve{04})$, $|\breve{08})\leftrightarrow|\breve{05})$\\
1.22        &   $|\breve{07})\leftrightarrow|\breve{19})$, $|\breve{08})\leftrightarrow|\breve{20})$\\
1.49        &   $|\breve{07})\leftrightarrow|\breve{21})$, $|\breve{08})\leftrightarrow|\breve{22})$\\
1.79        &   $|\breve{07})\leftrightarrow|\breve{01})$, $|\breve{08})\leftrightarrow|\breve{02})$\\
\hline
\end{tabular}
\caption{Identification of the peaks seen in the upper panel of Fig.\ \ref{fft-LL-RL-im-0G-Vgp0p50-dis1p0Em04-muL1p40-nrr0-hw2p00-gEM0px0}
         for the spectral density of the current-current correlations in the case of a $y$-polarized photon field. $g_\mathrm{EM}=0.05$ meV,
         $V_g=0.5$ mV, $\hbar\omega=2.00$ meV, and $\kappa = 10^{-4}$ meV.}
\label{Vg050-transitions}
\end{table}
The first and the third transitions in Table \ref{Vg050-transitions} are between initial and final states with
different mean photon number, radiative transitions.   
The energy of these transitions depends thus on the electron photon coupling,
$g_\mathrm{EM}$, but the second and the last four transitions are independent of this coupling as the photon component
of the final and initial states is low, mainly non-radiative transitions. This observation has to qualified
with the fact that states $|\breve{07})$ and  $|\breve{08})$ represent the lower branch of Rabi split pairs
that are quite detuned leaving a only a small photon component in them.

\begin{figure}[htb]
      \includegraphics[width=0.48\textwidth]{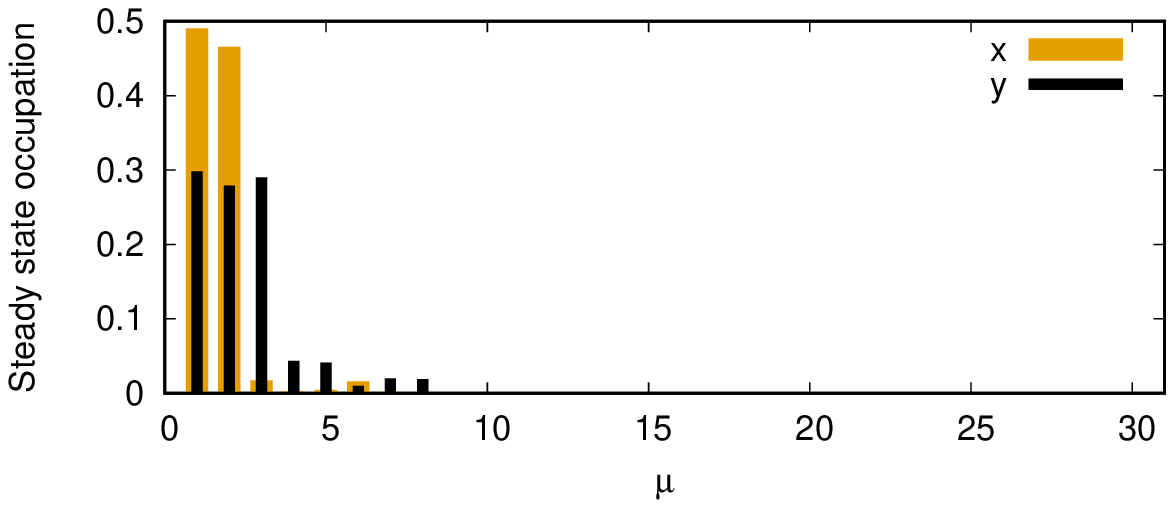}
      \includegraphics[width=0.48\textwidth]{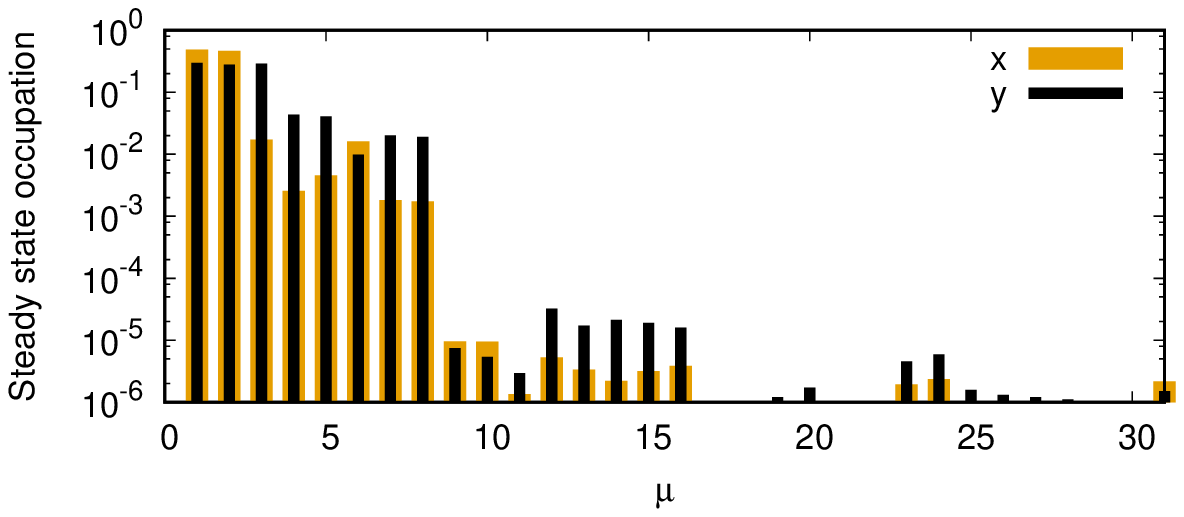}
      \caption{The steady state occupation of the central system on a linear (upper), and 
               logarithmic scale (lower) for the lowest 32 states, labeled with $\mu$, 
               and the polarization of the photon field indicated by the 
               width and the color of each column.
               $V_g=0.50$ mV, $\hbar\omega=2.00$ meV, and $\kappa = 10^{-4}$ meV.}
      \label{SS-occ-0G-Vgp0p50-dis1p0Em04-muL1p40-nrr0-hw2p00}
\end{figure}

The slight occupation of the one-electron states just above the bias window, $|\breve{07})$ and $|\breve{08})$, 
(see Fig.\ \ref{SS-occ-0G-Vgp0p50-dis1p0Em04-muL1p40-nrr0-hw2p00})
leads to strong Rabi-oscillations in the current correlations that is manifested by a dominant peak in its spectral 
density at the energy of the Rabi splitting, 0.2428 meV shown in Fig.\ \ref{fft-LL-RL-im-0G-Vgp0p50-dis1p0Em04-muL1p40-nrr0-hw2p00-gEM0px0}.
For $V_g=2.0$ mV, we had transitions from the one-electron ground state to the Rabi branches, but here at $V_g=0.5$ mV we observe 
an oscillation between the two Rabi states enabled by the special location of the states with respect to the 
bias window, even when the photon energy of the cavity, $\hbar\omega =2.0$ meV, is considerably detuned from the 
Rabi resonance at 1.8 meV. 
In Figs.\ \ref{fft-LL-RL-im-0G-Vgp0p50-dis1p0Em04-muL1p40-nrr0-hw2p00-gEM0px0} and
\ref{SS-occ-0G-Vgp0p50-dis1p0Em04-muL1p40-nrr0-hw2p00} we see transitions that lead
to the steady state occupation of both spin components of the one-electron ground
state, $|\breve{01})$ and $|\breve{02})$, and other states below the bias window.
In addition, we see a slight occupation of the lowest in energy spin-triplet two-electron
states, $|\breve{14})$, $|\breve{15})$ and $|\breve{16})$, especially for the $y$-polarized 
photon field. Only a tiny occupation of the Rabi split two-electron states $|\breve{23})$ and $|\breve{24})$ 
that are in resonance with the two-electron ground state $|\breve{06})$ can be seen in 
Fig.\ \ref{SS-occ-0G-Vgp0p50-dis1p0Em04-muL1p40-nrr0-hw2p00}.

\section{Summary}
We have used the current-current correlation spectral density of a multi-state
electron system placed in a photon cavity to calculate which transitions are 
active in the system in its steady state. The central system is weakly coupled to
the external leads, but in it the electrons couple strongly to the cavity photons.
In order to account for the influence of the geometry on the electron transport
through the system we have had to include the electron-electron Coulomb interaction
and both the electron-photon para- and diamagnetic interactions with numerical 
diagonalization in large many-body Fock spaces.\cite{Gudmundsson:2013.305} 
In a many-state system we find that it might be difficult to isolate individual resonances and 
thus we do not use the rotating wave approximation for the electron-photon interactions.

In order to effectively describe the transport in a system with diverse relaxation constants
through many orders of magnitude for the time variable we have mapped a non-Markovian
master equation into a Markovian master equation in 
Liouville space.\cite{Gudmundsson16:AdP_10,2016arXiv161003223J}
We have selected the decay constant of the photon cavity, $\kappa$,
(the coupling to the photon reservoir) to be of the same order of magnitude 
as the main relaxation channels of the electronic transitions to, or from, the
leads.

For the one-electron ground state inside the bias window defined by the two external
leads, we identify strong transitions to the Rabi split states of the first excitation
of the ground state for an appropriate photon energy. Several other weaker transitions are
seen in this case. In addition, we identify a transition to a higher order Rabi split state.  

For the two-electron ground state within the bias window and photon 
energy coupling it to its first excitation, we see a neighboring Rabi resonance for one-electron
states just above the bias window playing a strong role in the electron transport. 
In this case we also identify a transition between the two Rabi branches as they gain a
slight occupancy in the steady stead.

It is important to notice that in the noise spectral density for the current-current
correlation function we are able to identify both the radiative and the non-radiative many-body
transitions active in the system maintaining its steady state. The height of the spectral
peaks gives the weight or the strength of the different transitions, and their character, whether they
are radiative or not can be found by varying slightly the electron-photon coupling.
The peaks representing non-radiative transitions are stationary under that variation. 
The current noise power spectra are thus an important quantity to measure in experiments
on the systems to analyze their dynamics.

Our results point out, the importance of, and the opportunities in using the interplay
of geometry and photon polarization in transport of electrons through a nanoscale
electron system in a photon cavity. The double parallel quantum dots is the simplest
system offering clean separation of effects with its clear anisotropy.

\begin{acknowledgments}
This work was financially supported by the Research Fund of the University of Iceland,
the Icelandic Research Fund, grant no.\ 163082-051, 
and the Icelandic Instruments Fund. HSG and CST acknowledge support from Ministry of Science and 
Technology of Taiwan, under grant No.\ 103-2112-M-002 -003 -MY3, No.\ 106-2112-M-002 -013 -MY3, No.\
103-2112-M-239-001-MY3, and No.\ 106-2112-M-239-001-MY3.
\end{acknowledgments}

%
%
\bibliographystyle{apsrev4-1}
%

%
%
%
\end{document}